\documentclass[manuscript,screen,nonacm]{acmart}
\AtBeginDocument{%
  }

\usepackage{subcaption}
 \usepackage[ruled,vlined]{algorithm2e}

 \usepackage{multirow}
\usepackage{makecell}
\usepackage{graphicx} 
\usepackage{array}
\usepackage{multirow}
\usepackage{makecell}
\usepackage{arydshln}
\usepackage{booktabs}
\usepackage{rotating}     

\begin{document}

\title{Performance Optimization and Comparative Analysis of Generative AI Models on Advanced Accelerators}


\author{Amitash Nanda}
\authornote{This author contributed majorly to this research.}
\email{ananda@ucsd.edu}
\orcid{0000-0002-9792-2110}
\affiliation{%
  \institution{University of California San Diego}
  \city{San Diego}
  \state{CA}
  \country{USA}
}



\author{Javier Hernandez Nicolau}
\email{jhernandeznicolau@sdsc.edu}
\orcid{0000-0003-1470-1820}
\affiliation{%
  \institution{San Diego Supercomputer Center}
   \city{San Diego}
  \state{CA}
  \country{USA}
}

\author{Madhusudan Gujral}
\email{mgujral@ucsd.edu}
\orcid{0000-0001-5010-4826}
\affiliation{%
  \institution{San Diego Supercomputer Center}
   \city{San Diego}
  \state{CA}
  \country{USA}
}


\author{Mahidhar Tatineni}
\email{mahidhar@sdsc.edu}
\orcid{0009-0003-0709-090X}
\affiliation{%
  \institution{San Diego Supercomputer Center}
   \city{San Diego}
  \state{CA}
  \country{USA}
}

\author{Amitava Majumdar}
\email{majumdar@sdsc.edu}
\orcid{0000-0002-0860-6686}
\affiliation{%
  \institution{San Diego Supercomputer Center}
   \city{San Diego}
  \state{CA}
  \country{USA}
}
\author{Debashis Sahoo}
\email{dsahoo@ucsd.edu}
\orcid{0000-0003-2329-8228}
\affiliation{%
  \institution{University of California San Diego}
   \city{San Diego}
  \state{CA}
  \country{USA}
}
\renewcommand{\shortauthors}{nanda et al.}

\begin{abstract}
Generative AI models, such as Large Language Models (LLMs) and diffusion models, have demonstrated impressive performance across a wide range of tasks. Despite these advances, deployment remains challenging due to substantial memory requirements, extended inference latency, significant computational demands, and high hardware costs. These issues are further complicated when evaluating models across heterogeneous platforms, where differences in numerical formats, memory bandwidths, and software stacks interact with model architecture and workload characteristics in complex ways. To address these challenges, we present a systematic study focused on performance optimization and comparative analysis of several Generative AI models across diverse downstream tasks. This work introduces a novel mixed-precision post-training quantization evaluation, examines fine-tuning strategies, and assesses performance across modern high-performance computing (HPC) systems and advanced accelerators.  
\end{abstract}






\maketitle

\section{Introduction}

In recent years, generative AI models have been widely adopted by researchers, engineers, students, and others. Most are used for natural language processing, such as large language models (LLMs), or for image generation, such as diffusion models \cite{minaee2024large, cao2024survey, balija2024building}. Both approaches are based on traditional neural networks and transformer architecture, and have been heavily optimized, with core operations (mainly matrix multiplications) that can be parallelized and offloaded to specialized accelerators. These accelerators are available in a wide variety of architectures from leading vendors and differ significantly between generations. 
In this work, we present our analysis and evaluation using accelerators from three different high-performance computing (HPC) systems: Perlmutter~\cite{nersc2021perlmutter}, Expanse~\cite{expanse_pearc21}, and Voyager~\cite{voyager_pearc23}. Perlmutter is housed at NERSC, Lawrence Berkeley National Laboratory, while Expanse and Voyager are hosted at the San Diego Supercomputer Center (SDSC). Perlmutter is a heterogeneous system comprising both CPU- and GPU-accelerated nodes. It consists of $1,536$ GPU nodes, each with $1$ AMD Milan processor and $4$ NVIDIA A100 GPUs. Expanse is also a general-purpose, heterogeneous distributed compute cluster, organized into $13$ Scalable Compute Units (SSCUs), with $728$ standard AMD CPU nodes, $54$ NVIDIA V100 GPU nodes, $8$ NVIDIA A100 GPU nodes, $34$ NVIDIA H100 GPU nodes, and $4$ large-memory nodes. Voyager, on the other hand, is an innovative system specifically designed for AI applications, comprising $42$ nodes with Intel Gaudi accelerators. Currently, second-generation accelerators (Gaudi2) are replacing the original first-generation (Gaudi1) nodes. In addition, we were granted limited access to the third-generation Gaudi accelerator (Gaudi3) via the IBM cloud. The Gaudi architecture has evolved through three generations, with each adding more high-bandwidth memory (HBM) and tensor processing cores (TPCs) per HPU card. Specifically, Gaudi1 offers $32$ GB of HBM and $8$ TPCs, Gaudi2 provides $96$ GB of HBM and $24$ TPCs, and Gaudi3 delivers $128$ GB of HBM and $64$ TPCs. Each Gaudi node consists of $8$ HPU cards.

\section{Mixed-Precision Post Training Quantization on Voyager and Perlmutter}

Quantization for LLMs is crucial for the efficient deployment of large, resource-intensive models on real hardware. Post-training quantization (PTQ) reduces memory footprint and bandwidth usage without retraining. In LLMs, memory movement is often a bottleneck, so lowering precision can improve both capacity and throughput. PTQ methods help reduce memory bandwidth and accelerate inference. Different architectures can respond differently to quantization because of kernel design, scaling rules, and software maturity. Mixed-precision is useful because not all transformer submodules tolerate reduced precision equally well. Unlike full-model quantization, mixed-precision PTQs keep sensitive components at higher precision and apply lower precision to other parts of the model \cite{nanda2024cptquant}. In this work, we introduce a novel sensitivity-aware mixed-precision post-training quantization framework. This framework employs a two-phase approach to optimize LLM models across different accelerators (NVIDIA A100 GPUs and Intel Gaudi HPU). We evaluated the performance of the TinyLlama-1.1B-intermediate-step-1431k-3T base model on three diverse tasks: language modeling (WikiText-2), commonsense reasoning (HellaSwag), and reading comprehension (BoolQ). Our evaluation includes ``fake” quantization (simulated in FP32 for algorithm development) and ``real” quantization using FP16/INT8 for A100 and BF16/FP8 using Intel Neural Compressor for Gaudi accelerator. 

In the simulated quantization process, sensitivities are computed at the transformer layer level. The algorithm collects all $nn.Linear$ layer weights from the target layer and excludes biases, LayerNorm, and RMSNorm parameters. Further, it computes a unified pruning threshold by concatenating all weight magnitudes into a single vector and finding the k-th smallest absolute value, where k corresponds to the sparsity target ($0$-$100$\%). For our experiments, we have used $30$\% sparsity. This unified threshold ensures consistent sparsity across all linear modules within the layer. The pruning mask is then applied element-wise, setting weights below the threshold to zero while preserving all others. The target layer undergoes pruning while others remain completely unchanged. The sensitivity (S) is computed as $S_\ell = \text{PPL/Acc}_{\text{pruned},\ell} - \text{PPL/Acc}_{\text{baseline}}$. In the real quantization case, we use the same magnitude-pruning method on individual linear submodules. The candidate set contains the Llama-style projections \texttt{self\_attn.q\_proj}, \texttt{self\_attn.k\_proj}, \texttt{self\_attn.o\_proj}, \texttt{mlp.gate\_proj}, \texttt{mlp.up\_proj}, and \texttt{mlp.down\_proj}. The algorithm supports selecting a candidate type from these for sensitivity computation, and in this work, all computations are performed using only MLP submodules. This approach reveals fine-grained sensitivity patterns across submodules and, by pruning the same measures, individual impact on model performance. The sensitivity computation for all datasets across A100 and Gaudi2 follows the same pre-processing steps. We have used the validation split of WikiText-2, with samples of $512$ continuous tokens each, and evaluation uses a batch size of $4$. For Hellaswag, we used $2000$ samples from the training split with a batch size of $64$. Each sample requires scoring four candidate story endings using length-normalized log-likelihood, where the model computes the probability of each ending given the context. Similarly, for BoolQ, we sampled $2000$ from the training split with a batch size of $64$. Each sample constructs a prompt combining the passage and question, then scores both ``yes" and ``no" continuations by computing their log-likelihoods.
\begin{figure}[t]
  \centering
  \includegraphics[width=1.0\linewidth]{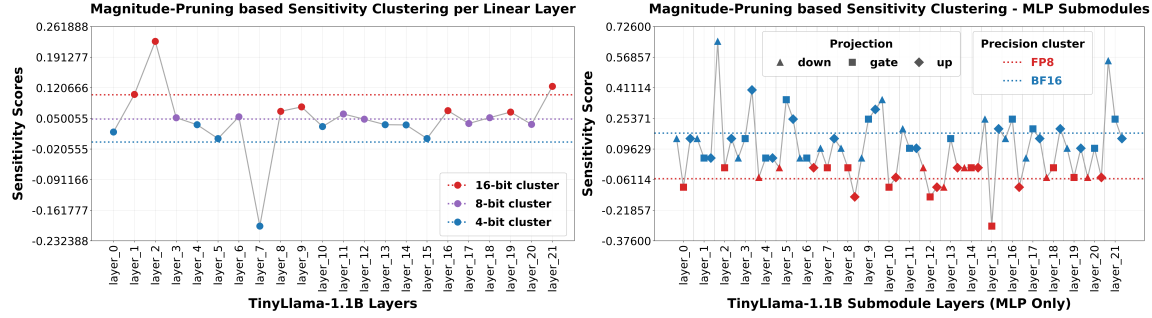}
  \caption{ Magnitude-pruning based sensitivity clustering of TinyLlama-1.1B linear layers for simulated quantization grouped into 16-8-4-bit precision cluster (left panel) and for submodules with FP8/BF16 real quantization (right panel).}
    \Description{The left panel of the figure shows a line graph plotted for all sensitivity values for each transformer linear layer. The sensitivity values are ranked and categorized into three clusters (16, 8, 4). The right panel shows the transformer, linear submodule sensitivity clustering, and precision assignment for FP8 and BF16 in real quantization.}
  \label{fig:clustering}
\end{figure}

The algorithm states that the layer's sensitivity to pruning is also affected by quantization. The sensitivity scores are clustered using a percentile method during the algorithm's evaluation phase, and the resulting cluster means are sorted from highest to lowest. These ranked clusters are mapped to a user-defined list of bit-widths for simulated quantization, such as [$16$, $8$, $4$], [$8$, $8$, $4$], or [$8$, $4$, $4$]. This produces a layer-to-bit map in which more sensitive layers receive higher precision and more robust layers receive lower precision as shown in Fig.~\ref{fig:clustering}. The evaluation phase uses group-wise quantization, dividing each weight matrix into groups of $128$ consecutive weights, computing separate scales and zero points for each group. The simulated quantization evaluation involves reshaping weights into groups, computing per-group statistics (min/max for asymmetric, max absolute value for symmetric), calculating quantization parameters, rounding to an integer grid, and immediately dequantizing back to FP32. WikiText-2 evaluation uses the entire test split and processes it using $512$-token continuous chunks, with perplexity as the primary metric. Hellaswag evaluation uses the full validation set of $10,042$ samples, and a maximum token length of $2048$. BoolQ evaluation also uses a full validation set of $3,270$ samples, with a maximum token length of 512. For both datasets, the algorithm uses accuracy as its primary metric. Gaudi’s real quantization leverages Intel Neural Compressor for hardware-accelerated FP8 operations. The implementation supports two FP8 formats: E4M3 offers better precision for small values, while E5M2 provides a wider dynamic range. The high-precision dtype defaults to BF16, which offers better training stability than FP16 while maintaining 16-bit storage. The quantization follows a two-pass approach: the MEASURE pass collects calibration statistics on representative data, while the QUANTIZE pass performs actual model conversion. Mixed-precision assignment uses sensitivity scores to allocate FP8 to robust layers and BF16 to sensitive layers. In real quantization evaluation on A100 GPU for high-precision layers, weights are stored as torch.float16 leveraging A100’s tensore cores for native FP16 accelration. While for aggressive quantization, weights are stored as torch.int8 with per-group FP16 scales. During forward passes, INT8 weights are dequantized to FP16 on the fly before matrix multiplication. Table~\ref {tab:quant-results} shows that evaluation of real quantization on A100 and Gaudi2 achieved 2-4x compression with minimal accuracy loss across three tasks.


\begin{table*}[t]
  \centering
  \caption{Mixed-Precision PTQ Performance Evaluation of TinyLlama-1.1B model on tasks: language modeling (WikiText-2), commonsense reasoning (HellaSwag), and reading comprehension (BoolQ) across A100 (1 GPU - 40 GB) and Gaudi2 (1 card - 80GB)}
  \label{tab:quant-results}
 
  \resizebox{\textwidth}{!}{%
    \renewcommand{\arraystretch}{1.25}
    \huge
    \begin{tabular}{@{} l l | c c | c c | c c | c c | c c @{}}
 
      \toprule
 
      \multirow{2}{*}{\textbf{Dataset}} &
      \multirow{2}{*}{\textbf{Metric}}  &
      \multicolumn{2}{c|}{\textbf{FP32}}             &
      \multicolumn{2}{c|}{\textbf{Fake [16, 8, 4]}}  &
      \multicolumn{2}{c|}{\textbf{Fake [8, 8, 4]}}   &
      \multicolumn{2}{c|}{\textbf{Fake [8, 4, 4]}}   &
      \multicolumn{2}{c}{\makecell{\textbf{Real Quantization}\\[-2pt]
                                   \textbf{[BF16, FP8] [FP16, INT8]}}} \\
 
      \cmidrule(lr){3-4}
      \cmidrule(lr){5-6}
      \cmidrule(lr){7-8}
      \cmidrule(lr){9-10}
      \cmidrule(lr){11-12}
 
      & &
      \textbf{Gaudi2} & \textbf{A100} &
      \textbf{Gaudi2} & \textbf{A100} &
      \textbf{Gaudi2} & \textbf{A100} &
      \textbf{Gaudi2} & \textbf{A100} &
      \textbf{Gaudi2} & \textbf{A100} \\
 
      \midrule
 
      \multirow{4}{*}{\textbf{Wikitext}}
        & Perplexity $\downarrow$
          & 12.02 & 12.02
          & 12.09 & 12.26
          & 12.09 & 12.80
          & 12.39 & 12.94
          & 12.02 & 12.02 \\
 
        & Compression Ratio $\uparrow$
          & 1x    & 1x
          & 3.47x & 3.47x
          & 4.80x & 4.80x
          & 5.93x & 5.93x
          & 2.42x & 3.77x \\
 
        & Throughput (tokens/sec) $\uparrow$
          & 32397 & 17024
          & 34235 & 17224
          & 33695 & 17203
          & 34081 & 17232
          & 23484 & 13547 \\
 
        & Evaluation Time (sec) $\downarrow$
          & 10.45 & 19.88
          &  9.89 & 19.65
          & 10.04 & 19.67
          &  9.93 & 19.64
          & 14.41 & 24.98 \\
 
      \midrule
  \specialrule{1.5pt}{0pt}{0pt}
      \multirow{4}{*}{\textbf{Hellaswag}}
        & Accuracy (\%) $\uparrow$
          & 56.86 & 56.97
          & 57.02 & 56.86
          & 57.04 & 56.70
          & 56.39 & 56.45
          & 56.92 & 56.92 \\
 
        & Compression Ratio $\uparrow$
          & 1x    & 1x
          & 3.28x & 3.47x
          & 4.67x & 4.80x
          & 5.74x & 5.93x
          & 2.3x  & 3.01x \\
 
        & Throughput (samples/sec) $\uparrow$
          & 22    & 13
          & 31    & 13
          & 32    & 13
          & 29    & 13
          & 14    & 50   \\
 
        & Evaluation Time (sec) $\downarrow$
          & 460.58 & 750.42
          & 325.62 & 748
          & 312.47 & 746
          & 351.42 & 745.14
          & 707    & 199.37 \\
 
      \midrule
  \specialrule{1.5pt}{0pt}{0pt}
      \multirow{4}{*}{\textbf{BoolQ}}
        & Accuracy (\%) $\uparrow$
          & 57.89 & 58.07
          & 54.43 & 54.31
          & 54.43 & 53.98
          & 52.91 & 51.87
          & 57.19   & 57.77 \\
 
        & Compression Ratio $\uparrow$
          & 1x    & 1x
          & 3.28x & 3.47x
          & 4.67x & 4.80x
          & 5.74x & 5.93x
          & 2.40x  & 3.01x \\
 
        & Throughput (samples/sec) $\uparrow$
          & 14    & 8
          & 22    & 9
          & 19    & 9
          & 24    & 9
          & 7   & 8    \\
 
        & Eval Time (sec) $\downarrow$
          & 236.43 & 410.69
          & 150.86 & 364.95
          & 174.61 & 367.83
          & 135.21 & 368.93
          & 453.31    & 405.95 \\
 
      \bottomrule
 
    \end{tabular}
}
\end{table*}

\section{Performance of fine-tuning LLM on Voyager}

Fine-tuning is the bridge between a general-purpose LLM and a highly specialized, up-to-date, and behaviorally tailored AI assistant. In this work, we assess the performance of fine-tuning tasks across different generations of Intel Habana Gaudi hardware, benchmarking with parallelization strategies using varying numbers of HPU cards. We selected three open-source LLMs from the Hugging Face Hub for this evaluation: Llama3.1:8B-Instruct (small), Gemma2:27B-It (medium), and Llama3.3:70B-Instruct (large). For the fine-tuning task, we utilized a psychology dataset (BoltMonkey) from the Hugging Face Hub, consisting of approximately 392,000 prompt-response records. 

\begin{figure}[t]
  \centering
  \includegraphics[width=1.0\linewidth]{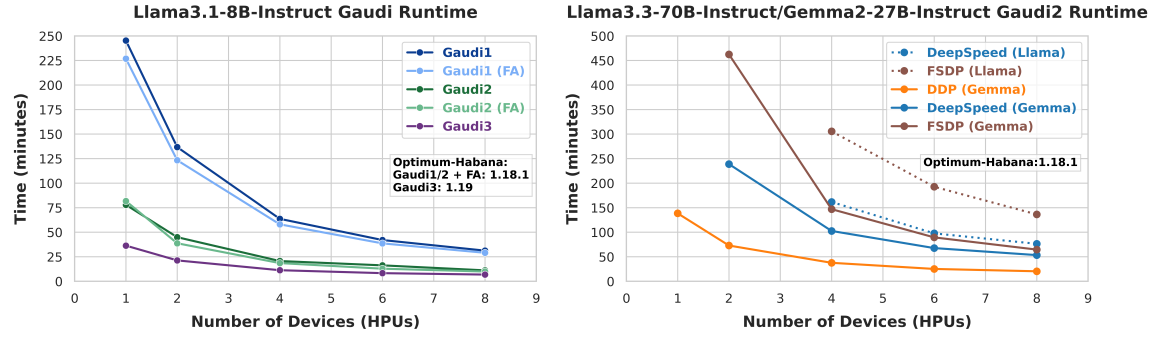}
  \caption{Fine-tuning runtime of the Llama3.1-8B-Instruct model across three generations of Gaudi nodes; FA refers to flash attention (left panel). Fine-tuning results for the Gemma2:27B-Instruct and Llama3.3:70B-Instruct models on the Gaudi2 node (right panel). }
\Description{The left panel graph is a line graph showing the fine-tuning runtime of the Llama3.1-8B-Instruct model across three generations of Gaudi nodes, with varying numbers of HPU devices. The y-axis represents the time taken for the evaluation. The right panel illustrates the fine-tuning results of a medium and a large model on the Gaudi2 node for Gemma2:27B-Instruct (solid lines) and Llama3.3:70B-Instruct (dotted lines), respectively.}
  \setlength{\belowcaptionskip}{-10pt} 

  \label{fig:fine-tuning}
\end{figure}

For our experiments, we utilized the Optimum-Habana library and configured a set of hyperparameters for a single epoch. A Gaudi1 card with a limited HBM can only accommodate a smaller LLM like Llama3.1:8B-Instruct. Employing more HPU cards with MPI-based Distributed Data Parallel (DDP) does not mitigate this memory limitation, as DDP replicates model parameters on each card while only sharding the input data. As shown in the left panel of Fig.~\ref{fig:fine-tuning}, the fine-tuning performance on Gaudi1 scales nearly linearly with the number of HPU cards. A similar pattern is observed with Gaudi2, but these cards are significantly faster, offering a substantial reduction in runtime of 2.5 to over 3 times faster than Gaudi1, depending on the number of cards provided. This improvement is attributable to the 24 tensor processing cores (TPCs) on each Gaudi2 card, triple the number found on Gaudi1.
Gaudi3 cards are also considerably faster than Gaudi2, demonstrating a performance gain of 1.8 to over 2-fold. Furthermore, utilizing flash attention (FA), natively supported in PyTorch, on Gaudi1 and Gaudi2 nodes resulted in noticeable performance improvements. FA is an optimized attention algorithm that minimizes data movement between different types of HPU memory, preventing bottlenecks by processing attention in smaller, SRAM-resident tiles of the Query (Q), Key (K), and Value (V) matrices. For most runs on Gaudi1 and Gaudi2, FA provided a performance gain of 10\% to 20\%.

Since larger models often cannot be accommodated on the available VRAM of a single Gaudi card, methodologies to alleviate this limitation are necessary. For example, the Llama3.3-70B-Instruct model cannot reside in the memory of Gaudi1, Gaudi2, or Gaudi3 cards. Therefore, fine-tuning larger models requires sharding mechanisms, such as the ZeRO3 implementation of DeepSpeed or fully sharded data parallel (FSDP). DeepSpeed offers three distinct sharding schemes, with ZeRO3 being the most advanced. It shards optimizer states, gradients, and the model, thereby substantially reducing memory usage during fine-tuning. However, as shown in the right panel of Fig.~\ref{fig:fine-tuning}, even with DeepSpeed, at least 4 Gaudi2 cards are required to fine-tune a 70B-parameter model. The alternative sharding approach, FSDP, is natively supported in PyTorch and requires no external libraries. While both DeepSpeed and FSDP utilize various sharding options, our experience on Voyager consistently demonstrates that DeepSpeed outperforms FSDP.
Unlike the 70B parameter model, Gemma2:27B-Instruct, a medium-sized model, can be accommodated on a single Gaudi2 card. The right panel of Fig.~\ref{fig:fine-tuning} shows the fine-tuning performance of this 27B-parameter model with DDP, DeepSpeed, and FSDP. These results clearly indicate that DDP decisively outperforms any model sharding procedure. Although DeepSpeed and FSDP reduce memory footprint through sharding, this introduces significant overhead and degrades performance. Our evaluation explicitly demonstrates that DeepSpeed or FSDP should be used only when a model cannot fit within the HBM of a single Gaudi card.


\begin{figure}[t]
  \centering
  \includegraphics[width=1.0\linewidth]{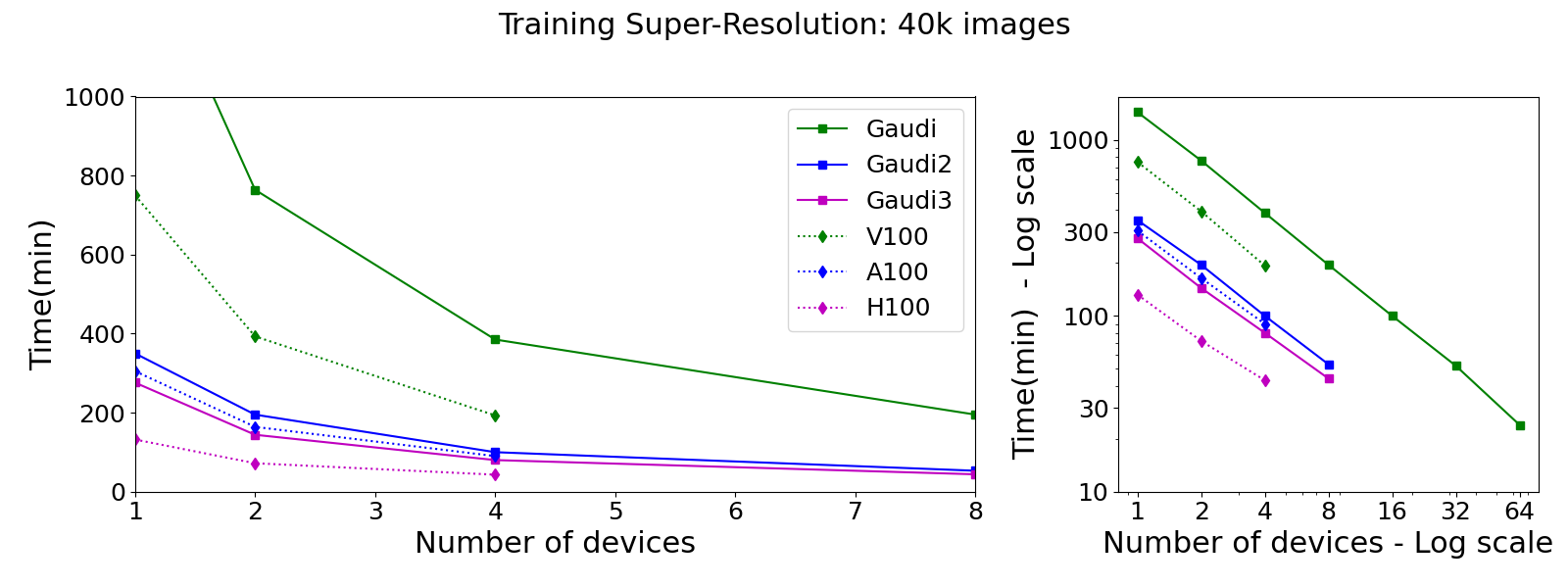}
  \caption{Super resolution model training time for different accelerators.}
  \Description{Two scatter plots showing the training time of a Diffusion model on different accelerators.}
  \label{fig:super-resolution}
\end{figure}
\section{Performance of Diffusion models on Voyager}

A Super-Resolution model~\cite{saharia2021_SR3,sr3_repo} has been used as a test base to analyze the performance of diffusion models on the Gaudi nodes. Astrophysicists are using this AI model to enhance the resolution of images of galactic dust emissions generated by PySM~\cite{Panexp_2025, Zonca_2021}. This model, based on the diffusion algorithm, has been ported to the Gaudi architecture and parallelized using PyTorch DDP. We have executed the same task (training with over 40,000 images) on different generations of the Gaudi accelerators. The model runs on SynapseAI version 1.21.4 on Lazy mode. A diffusion-based model cannot run in Eager mode on the Gaudi accelerator, which is the default and optimized mode. As a baseline, the same task was executed on different NVIDIA GPUs (V100, A100, H100) on Expanse. The training results are illustrated in Fig.~\ref{fig:super-resolution}. The left plot shows the analysis of up to 8 accelerators; Gaudi and NVIDIA GPU results are depicted in solid and dashed lines, respectively. For the GPUs, the model exhibits a speed-up of almost 2 from the V100 (green) to the A100 (blue), and another 2-fold gain from the A100 (blue) to the H100 (magenta). For Gaudi architecture, a 4x speed-up is observed from Gaudi1 to Gaudi2; however, this trend does not continue from Gaudi2 to Gaudi3 (solid blue and magenta lines). The right plot shows the same data on a log scale on both axes, highlighting near-perfect scaling.

\begin{acks}
This work was supported by the U.S. NSF Award 1928224 and 2005369, and allocation awards of ACCESS CIS251090, NAIRR260091, and DOE NERSC ASCR ERCAP0037521.
\end{acks}

\bibliographystyle{ACM-Reference-Format}
\bibliography{sample-base}
\appendix
\section{Reproducibility Artifact}
The code for this work can be found at \url{https://github.com/amitashnanda/Voyager_Gaudi_Benchmarks.git}



\end{document}